\documentstyle[12pt,epsfig]{article}
\newcommand{\agt}{\,\rlap{\lower 3.5 pt \hbox{$\mathchar \sim$}} \raise 1pt
 \hbox {$>$}\,}
\newcommand{\alt}{\,\rlap{\lower 3.5 pt \hbox{$\mathchar \sim$}} \raise 1pt
 \hbox {$<$}\,}
\catcode`@=11
\newcount\@tempcntc
\def\@citex[#1]#2{\if@filesw\immediate\write\@auxout{\string\citation{#2}}\fi
  \@tempcnta\z@\@tempcntb\m@ne\def\@citea{}\@cite{\@for\@citeb:=#2\do
    {\@ifundefined
       {b@\@citeb}{\@citeo\@tempcntb\m@ne\@citea\def\@citea{,}{\bf ?}\@warning
       {Citation `\@citeb' on page \thepage \space undefined}}%
    {\setbox\z@\hbox{\global\@tempcntc0\csname b@\@citeb\endcsname\relax}%
     \ifnum\@tempcntc=\z@ \@citeo\@tempcntb\m@ne
       \@citea\def\@citea{,}\hbox{\csname b@\@citeb\endcsname}%
     \else
      \advance\@tempcntb\@ne
      \ifnum\@tempcntb=\@tempcntc
      \else\advance\@tempcntb\m@ne\@citeo
      \@tempcnta\@tempcntc\@tempcntb\@tempcntc\fi\fi}}\@citeo}{#1}}
\def\@citeo{\ifnum\@tempcnta>\@tempcntb\else\@citea\def\@citea{,}%
  \ifnum\@tempcnta=\@tempcntb\the\@tempcnta\else
   {\advance\@tempcnta\@ne\ifnum\@tempcnta=\@tempcntb \else \def\@citea{--}\fi
    \advance\@tempcnta\m@ne\the\@tempcnta\@citea\the\@tempcntb}\fi\fi}
\catcode`@=12
\begin{document}
\title{\vskip-3cm{\baselineskip14pt
\centerline{\normalsize DESY 97--241\hfill ISSN~0418--9833}
\centerline{\normalsize MPI/PhT/97--88\hfill}
\centerline{\normalsize hep--ph/9712482\hfill}
\centerline{\normalsize December 1997\hfill}}
\vskip1.5cm
Predictions for $D^{*\pm}$ Photoproduction at HERA with New Fragmentation
Functions from LEP1}
\author{J. Binnewies$^1$, B.A. Kniehl$^2$, G. Kramer$^1$\\
$^1$ II. Institut f\"ur Theoretische Physik\thanks{Supported
by Bundesministerium f\"ur Bildung und Forschung, Bonn, Germany, under
Contract 05~7~HH~92P~(0), and by EU Program {\it Human Capital and Mobility}
through Network {\it Physics at High Energy Colliders} under Contract
CHRX--CT93--0357 (DG12 COMA).},
Universit\"at Hamburg,\\
Luruper Chaussee 149, 22761 Hamburg, Germany\\
$^2$ Max-Planck-Institut f\"ur Physik (Werner-Heisenberg-Institut),\\
F\"ohringer Ring 6, 80805 Munich, Germany}
\date{}
\maketitle
\begin{abstract}
We present new sets of nonperturbative fragmentation functions for $D^{*\pm}$
mesons, both at leading and next-to-leading order in the $\overline{\rm MS}$
factorization scheme with five massless quark flavors.
They are determined by fitting the latest OPAL and ALEPH data on inclusive
$D^{*\pm}$ production in $e^+e^-$ annihilation.  
We take the charm-quark fragmentation function to be of the form proposed by
Peterson et al.\ and thus obtain new values of the $\epsilon_c$ parameter,
which are specific for our choice of factorization scheme.
With these fragmentation functions, recent data on inclusive $D^{*\pm}$
photoproduction in $ep$ collisions at HERA are reasonably well reproduced.

\end{abstract}
\newpage

\section{Introduction}

Recently, the H1 \cite{1} and ZEUS \cite{2} collaborations at HERA presented
data on the differential cross section $d^2\sigma/dy_{\rm lab}\,dp_T$ of
inclusive $D^{*\pm}$ production in low-$Q^2$ $ep$ collisions, equivalent to 
photoproduction.
Here $y_{\rm lab}$ and $p_T$ are the rapidity and transverse momentum of the
produced $D^{*\pm}$ mesons in the laboratory frame, respectively. 
These measurements extended up to $p_T=12$~GeV.
Another source of information on $D^{*\pm}$ production comes from
$e^+e^-\to D^{*\pm}+X$ at the $Z$-boson resonance \cite{3,4}.
In this process, two mechanisms of $D^{*\pm}$ production contribute with
similar rates.
The $D^{*\pm}$ mesons are either produced by $Z\to c\bar c$ decay and
subsequent $c/\bar c\to D^{*\pm}$ fragmentation, or by $Z\to b\bar b$ decay
with subsequent fragmentation of $b/\bar b$ quarks into $B$ mesons, which
weakly decay into $D^{*\pm}$ mesons.
The latter two-step process is usually treated as a one-step fragmentation
process $b/\bar b\to D^{*\pm}$.
With the aid of very efficient bottom-tagging methods, these two main sources
of inclusive $D^{*\pm}$ production were disentangled with high purity, and
separate cross sections for the two production mechanisms were presented.
This allows one to determine separate fragmentation functions (FF's) for
$c/\bar c\to D^{*\pm}$ and $b/\bar b\to D^{*\pm}$.
Owing to the factorization theorem, these can then be used to make 
quantitative predictions for $D^{*\pm}$ production in other reactions such as
$\gamma p\to D^{*\pm}+X$, which is being measured at HERA.
Such a program, which constitutes a test of the universality of the $D^{*\pm}$
FF's, was recently carried out by two groups \cite{5,6}. 

These works are based on the so-called massless-charm scheme \cite{7}.
In this scheme, the charm-quark mass $m_c$ is neglected, except in the initial
conditions for the parton density functions (PDF's) and the FF's of the charm 
quarks.
This should be a reasonable approximation for center-of-mass (CM) energies
$\sqrt s\gg m_c$ in $e^+e^-$ annihilation or transverse momenta $p_T\gg m_c$
in $\gamma p$ scattering.
In this approach, charm is considered to be one of the active flavors inside
the initial photons and protons, in the same way as the lighter $u$, $d$, and
$s$ quarks.
Then, the collinear singularities which correspond to the
$\alpha_s\ln(\mu^2/m_c^2)$ terms, where $\mu$ is an appropriate factorization
scale ($\mu=\sqrt s$ and $\mu=p_T$, respectively), in a scheme where the charm
quark is treated as a massive particle and only three active flavors are
taken into account are absorbed into the charm-quark PDF's and FF's.
Instead of absorbing the logarithmic terms, one can start with $m_c=0$ and
absorb the initial- and final-state collinear singularities as usual in the
modified minimal-subtraction ($\overline{\rm MS}$) factorization scheme, based
on dimensional regularization.
These two possibilities to achieve a massless-charm result, {\it i.e.}, to
absorb the collinear singularities either in the form of logarithms or in the
form of $1/\epsilon$ poles via $\overline{\rm MS}$ factorization, differ by
finite terms, as was shown in Ref.~\cite{8} for the case of
$e^+e^-\to c/\bar c+X$.
These terms can be considered as perturbative FF's at the low initial scale
$\mu_0$ of order $m_c$, which are evolved to higher scales with the usual
Altarelli-Parisi (AP) equations \cite{9}.
Along these lines, cross sections for $p\bar p$ \cite{10}, $\gamma p$
\cite{11}, and $\gamma\gamma $ \cite{12} collisions were calculated in
next-to-leading order (NLO), describing this way the transformation of a
massless charm quark into a massive charm quark via a perturbative QCD cascade.
Later, this approach was extended by including a nonperturbative FF which
describes the transition from the massive charm quark and antiquark to the
$D^{*\pm}$ mesons \cite{5,13}.
In our earlier works on $D^{*\pm}$ production \cite{6,14}, instead of working
with the perturbative FF's, we incorporated their effect by an equivalent
modification of the pure $\overline{\rm MS}$ scheme with massless quarks,
which led to a particular kind of massive subtraction scheme.
This change of scheme was restricted to the final state.

After fitting the nonperturbative fragmentation component to the $e^+e^-$
data taken by ARGUS \cite{15} and OPAL \cite{4} in Ref.~\cite{5} and to those
taken by ALEPH \cite{3} and OPAL \cite{4} in Ref.~\cite{6}, the cross section
of inclusive $D^{*\pm}$ photoproduction in $ep$ collisions as measured by H1
\cite{1} and ZEUS \cite{2} could be reasonably well described.
In particular, not only the shape, but also the normalization of the cross
section was very well accounted for \cite{6}.
These papers showed that, in order to successfully predict inclusive
$D^{*\pm}$ production in low-$Q^2$ $ep$ collisions, it is essential to
incorporate information on the fragmentation process from other reactions and
to use the very same factorization scheme for all considered processes.

The fragmentation of charm or bottom quarks into $D^{*\pm}$ mesons cannot be
calculated from first principles in perturbative QCD.
In fact, in order to realistically describe the formation of the $D^{*\pm}$
mesons, a nonperturbative component, which is not known theoretically, is
always needed.
Hence, it is certainly appropriate to give up the perturbative component of
the FF input altogether and to describe the $c/\bar c\to D^{*\pm}$ and
$b/\bar b\to D^{*\pm}$ transitions entirely by nonperturbative FF's, as is
usually done for the fragmentation of $u$, $d$, and $s$ quarks into light
mesons.

The aim of the present work is to reconsider the production of $D^{*\pm}$
mesons both in $e^+e^-$ annihilation and photoproduction adopting this
puristic approach of fully nonperturbative fragmentation in the
$\overline{\rm MS}$ scheme with five massless flavors.
This analysis is based on very recent high-statistics data from the OPAL
\cite{16} and ALEPH \cite{17} collaborations at LEP1.
The FF's thus fitted are then used to update our predictions for the cross
section of $D^{*\pm}$ photoproduction in $ep$ collisions to be compared with
new data from the ZEUS collaboration \cite{18}.

The outline of this paper is as follows.
In Sec.~2, we briefly recall the theoretical framework underlying the
extraction of FF's from $e^+e^-$ data, which has already been introduced in
Refs.~\cite{6,14}.
Then, we present the FF's we obtained by fitting the new $D^{*\pm}$ data from
OPAL \cite{16} and ALEPH \cite{17} at leading order (LO) and NLO in the pure
$\overline{\rm MS}$ factorization scheme with five massless flavors.
In Sec.~3, we apply the nonperturbative parameters determined in Sec.~2 to 
make LO and NLO predictions for $D^{*\pm}$ photoproduction in $ep$ collisions,
which we then compare with the latest ZEUS data \cite{18}.
Our conclusions are summarized in Sec.~4.

\boldmath
\section{$D^{*\pm}$ Production in $e^+e^-$ Annihilation}
\unboldmath

Our procedure to construct LO and NLO sets of $D^{*\pm}$ FF's has already been
described in our previous paper \cite{6}.
Here, we only give those details which differ from Ref.~\cite{6}.
As experimental input, we now use the new LEP1 data from OPAL \cite{16} and
ALEPH \cite{17}.
While the older ALEPH \cite{3} and OPAL data \cite{4} agreed well enough to be
simultaneously fitted \cite{6}, the new data samples do not sufficiently
overlap.
Thus, we refrain from performing a combined fit.
Instead, we generate independent LO and NLO FF sets for the two data samples.
This allows us to check actually how well our predictions for other
$D^{*\pm}$-production cross sections are constrained by the available
information from $e^+e^-$ annihilation, an error source which has so far been
rather difficult to assess quantitatively.

In $e^+e^-$ annihilation at the $Z$-boson resonance, charmed mesons are
produced either directly through the hadronization of charm quarks produced 
by $Z\to c\bar c$ or via the weak decays of $B$ hadrons from $Z\to b\bar b$,
with an approximately equal rate. 
Charmed mesons from $Z\to c\bar c$ allow us to determine the charm-quark FF.
The main achievement of both experimental analyses \cite{16,17} was to
disentangle these two main sources of $D^{*\pm}$ production at the $Z$-boson
resonance.
In Ref.~\cite{16}, a combination of several charm-quark tagging methods, based
on fully and partially reconstructed $D^{*\pm}$ mesons, and a bottom-quark 
tag, based on identified electrons and muons, was used.
In Ref.~\cite{17}, the separation of $c\bar c$ and $b\bar b$ events was
improved by means of an algorithm based on the measurement of the impact
parameter of charged tracks and the effective mass of those least fitted to
the primary vertex.
As in the earlier analyses \cite{3,4}, the $D^{*\pm}$ mesons were identified
via the decay chain $D^{*+}\to D^0\pi^+$ and $D^0\to K^-\pi^+$ (and the
analogous chain for the $D^{*-}$ meson), which allows for a particularly clean
signal reconstruction.
The experimental cross sections \cite{16,17} were presented as distributions
differential in $x=2E(D^{*\pm})/\sqrt s$, where $E(D^{*\pm})$ is the measured
energy of the $D^{*\pm}$ candidate, which are normalized to the total number
of multihadronic $Z$-boson decays.
Besides the total $D^{*\pm}$ yield, which receives contributions from
$Z\to c\bar c$ and $Z\to b\bar b$ decays as well as from gluon fragmentation,
both experimental groups separately specified results for $D^{*\pm}$ mesons
from tagged $Z\to b\bar b$ events.
The contribution due to charm-quark fragmentation is peaked at large $x$,
whereas the one due to bottom-quark fragmentation has its maximum at small
$x$.

For the fits, we use the $x$ bins in the interval $[0.1,1.0]$ and integrate
the theoretical functions over the bin widths, which is equivalent to the
experimental binning procedure.
As in the experimental analyses, we sum over $D^{*+}$ and $D^{*-}$ mesons.
As a consequence, there is no difference between the FF's of a given quark
and its antiquark.
As in Ref.~\cite{6}, we take the starting scales for the $D^{*\pm}$ FF's of
the gluon and the $u$, $d$, $s$, and $c$ quarks and antiquarks to be
$\mu_0=2m_c$, while we take $\mu_0=2m_b$ for the FF's of the $b$ quark and
antiquark.
The FF's of the gluon and the first three flavors are assumed to be zero at
the starting scale.
These FF's are generated through the $\mu^2$ evolution, and the FF's of the
first three quarks and antiquarks coincide with each other at all scales $\mu$.

We employ two different forms for the parameterization of the charm- and
bottom-quark FF's at their respective starting scales.
In the case of charm, we use the Peterson distribution \cite{19},
\begin{equation}
\label{peterson}
D_c(x,\mu_0^2)=N\frac{x(1-x)^2}{[(1-x)^2+\epsilon_c x]^2}.
\end{equation}
In the case of bottom, we adopt the ansatz
\begin{equation}
\label{standard}
D_b(x,\mu_0^2)=Nx^{\alpha}(1-x)^{\beta},
\end{equation}
which is frequently used for the FF's of light hadrons.
However, in contrast to the light-hadron case, we expect that $\alpha>0$ for
$D^{*\pm}$ mesons, so that $D_b(x,\mu_0^2)\to0$ as $x\to0$.
Furthermore, we expect that $\alpha<\beta$, since $D_b(x,\mu_0^2)$ is supposed
to have its maximum somewhere at $x<0.5$.
This choice of starting distributions was also used for the mixed set (M) in
Ref.~\cite{6}.
The Peterson form is particularly suitable for FF's that peak at large $x$.
Since the bottom-quark FF, being a convolution of the $b\to B$ fragmentation
and the subsequent $B\to D^{*\pm}+X$ decay, has its maximum at small $x$
values, we obtain intolerably bad fits if we also use Eq.~(\ref{peterson}) in
this case.
In Ref.~\cite{6}, we also studied the alternative where Eq.~(\ref{standard})
is used both for the charm and bottom quarks, leading to the standard set (S).
Although set S yielded slightly lower $\chi^2$ values per degree of freedom
($\chi_{\rm DF}^2$) at the expense of having one more fit parameter, both
sets, S and M, led to almost identical predictions for $D^{*\pm}$
photoproduction in the kinematical regions of present interest at HERA.
In this work, we thus limit ourselves to the mixed ansatz.
This facilitates the comparison of our results, especially on $\epsilon_c$,
with the literature on charm-quark fragmentation, where the Peterson
ansatz~(\ref{peterson}) is commonly used.
Ansatz~(\ref{standard}) was also employed some time ago in Ref.~\cite{13} to
describe the nonperturbative FF's of charm quarks into $D$ and $D^*$ mesons
and those of bottom quarks into $B$ mesons.

The calculation of the cross section $(1/\sigma_{\rm tot})d\sigma/dx$ for
$e^+e^-\to\gamma,Z\to D^{*\pm}+X$ is performed as described in Ref.~\cite{6},
except that we now abandon the subtraction terms $d_{Qa}(x)$ specified in
Eq.~(7) therein, for reasons explained in Sec.~1.
All relevant formulas and references may be found in Ref.~\cite{6}.

Both OPAL \cite{16} and ALEPH \cite{17} presented momentum distributions for
their full $D^{*\pm}$ samples and for their $Z\to b\bar b$ subsamples.
We received these data in numerical form via private communications
\cite{16,17}.
The OPAL data are displayed in Fig.~4 of Ref.~\cite{16} in the form
$(1/N_{\rm had})dN_{D^{*\pm}}/dx$, where $N_{D^{*\pm}}$ is the number of
$D^{*\pm}$ candidates reconstructed through the decay chain mentioned above.
In order to convert this into the cross section
$(1/\sigma_{\rm tot})d\sigma/dx$, we need to divide by the branching 
fractions $B(D^{*+}\to D^0\pi^+)=0.683\pm0.014$ and
$B(D^0\to K^-\pi^+)=0.0383\pm0.0012$ \cite{20}.
The momentum distribution of the full ALEPH sample, which is shown in Fig.~3
of Ref.~\cite{17}, has to be treated in the same way.
In the case of the ALEPH $Z\to b\bar b$ subsample, which is presented in
Fig.~8 of Ref.~\cite{17}, we need to include the additional factor
$R_b\times f_b=0.0464$, where
$R_b=\Gamma(Z\to b\bar b)/\Gamma(Z\to {\rm hadrons})$ and $f_b$ is the
fraction of $b\bar b$ events in the sample which remains after subtracting the
events from gluon splitting; this factor may be extracted from Ref.~\cite{17}.
The ALEPH data \cite{17} have slightly smaller statistical errors than the
OPAL data \cite{16}.

As for the asymptotic scale parameter appropriate for five active quark
flavors, we adopt the LO (NLO) value
$\Lambda_{\overline{\rm MS}}^{(5)}=108$~MeV (227~MeV) from our study of
inclusive charged-pion and -kaon production \cite{21}.
The particular choice of $\Lambda_{\overline{\rm MS}}^{(5)}$ is not essential;
other values, in particular higher LO values of
$\Lambda_{\overline{\rm MS}}^{(5)}$, can easily be accommodated by slightly
shifting the other fit parameters without changing the quality of the fit.
As in Ref.~\cite{6}, we take the charm- and bottom-quark masses to be
$m_c=1.5$~GeV and $m_b=5$~GeV, respectively.
Since $m_c$ and $m_b$ only enter via the definitions of the starting scales
$\mu_0$ of the FF's, their precise values are immaterial for our fit.

The values of $N$ and $\epsilon_c$ in Eq.~(\ref{peterson}) and of $N$,
$\alpha$, and $\beta$ in Eq.~(\ref{standard}) which result from our LO and NLO
fits to the OPAL and ALEPH data are summarized in Table~\ref{pars}.
In the following, we refer to the corresponding FF's as sets LO~O, NLO~O,
LO~A, and NLO~A, respectively.
In Table~\ref{fit}, we list three $\chi_{\rm DF}^2$ values for each of these
four fits: one for the $Z\to b\bar b$ subsample; one for the total sample (sum
of $c\bar c$-tagged, $b\bar b$-tagged, and gluon-splitting events); and the
average evaluated by taking into account both the $Z\to b\bar b$ subsample and
the total sample.
We observe that all four fits are quite successful, with $\chi_{\rm DF}^2$
values of order unity.
The LO and NLO fits to the OPAL data are slightly better than those to the
ALEPH data.
In each case, the $Z\to b\bar b$ subsample tends to be less well described by
the fit than the total sample.
As expected on general grounds, the NLO fits are superior to the LO fits.

In Fig.~\ref{ee}a, we compare the OPAL data \cite{16} with the LO and NLO
calculations using sets LO~O and NLO~O, respectively.
The analogous analysis for the ALEPH data \cite{17} is shown in
Fig.~\ref{ee}b.
Notice that the distributions plotted in Figs.~\ref{ee}a and b correspond to
$(1/\sigma_{\rm tot})d\sigma/dx$, {\it i.e.}, the experimental data are
multiplied by the factors specified above.
Except at very small $x$, the LO and NLO results are very similar. 
This is also true for the distributions at the starting scale, as may be seen
by comparing the corresponding LO and NLO parameters in Table~\ref{pars}.
Only $\epsilon_c$ changes appreciably as we pass from LO to NLO.
The branching of the LO and NLO results at small $x$ indicates that, in this
region, the perturbative treatment ceases to be valid. 
This is related to the phase-space boundary for the production of $D^{*\pm}$
mesons at $x_{\rm min}=2m(D^{*\pm})/\sqrt s$, where $m(D^{*\pm})$ is the
$D^{*\pm}$ mass.
At $\sqrt s=M_Z$, one has $x_{\rm min}=0.046$.
This is approximately where our NLO results turn negative.
Since our massless-quark approach is not expected to be valid in regions of
phase space where finite-$m(D^{*\pm})$ effects are important, our results 
should only be considered meaningful for $x\agt x_{\rm cut}=0.1$, say.
We also encountered a similar small-$x$ behaviour in our previous analysis
\cite{6}, where we fitted the older ALEPH \cite{3} and OPAL \cite{4} data in
the framework of the massive subtraction scheme.
Since the two rightmost data points of the ALEPH $Z\to b\bar b$ sample,
corresponding to the $x$ bins $[0.85,0.9]$ and $[0.9,0.95]$, come with
negative cross sections and rather small errors \cite{17}, we excluded them
from our fits.

Comparing Figs.~\ref{ee}a and b, we observe that the OPAL and ALEPH data are
indeed somewhat different in shape and do not mutually overlap within their
errors.
Thus, combined LO and NLO fits would lead to intolerable values of
$\chi_{\rm DF}$.
By the same token, the separate fits to the OPAL and ALEPH data lead to 
significantly different values for the parameters appearing in the starting
distributions (\ref{peterson}) and (\ref{standard}), as may be seen from
Table~\ref{pars}.
There are also striking differences between the parameters resulting from the
LO and NLO fits to the same data sets.
The value $\epsilon_c=0.0851$ of set LO~O is very similar to the value
$\epsilon_c=0.0856$ of set LO~M in our previous analysis \cite{6}.
On the other hand, our new NLO results for $\epsilon_c$, namely 0.116 for set
NLO~O and 0.185 for set NLO~A, are significantly larger than our previous
value 0.0204 for set NLO~M \cite{6}.
This dramatic shift in $\epsilon_c$ is due to the fact that we are now using a
different scheme for the factorization of the final-state collinear 
singularities (namely, the pure $\overline{\rm MS}$ scheme with five massless
flavors) than in Ref.~\cite{6}, where we modified this scheme so as to
incorporate the effect of the perturbative FF's \cite{8} (massive subtraction
scheme).
One important advantage of our new approach becomes apparent if we compare
Figs.~\ref{ee}a and b with Figs.~1b and 2b in Ref.~\cite{6}; see also
Fig.~1 in Ref.~\cite{5}.
While in the NLO analyses of Refs.~\cite{5,6}, the NLO cross section of
$e^+e^-\to D^{*\pm}+X$ turned negative for $x\agt0.9$, the new NLO 
calculation stays positive in the upper $x$ range, coincides there with the LO
calculation, and nicely describes the data.
We attribute the unphysical large-$x$ behaviour of the NLO calculation in 
Refs.~\cite{5,6} to the use of the massive subtraction scheme in conjunction
with the special forms for the nonperturbative FF's at the starting scale.
This led to the appearance of large Sudakov logarithms which spoiled the NLO
result at $x\agt0.9$.
This feature forced us \cite{6}, and also the authors of Ref.~\cite{5}, to
exclude from the fits the rather precise experimental data points at $x$ close
to unity.
By contrast, the NLO calculation in the pure $\overline{\rm MS}$ scheme does 
not suffer from theoretical problems at high $x$, and there is no need to omit
valuable experimental information in that region.

As mentioned above, we take the FF's of the partons
$g,u,\bar u,d,\bar d,s,\bar s$ to be vanishing at their starting scale
$\mu_0=2m_c$.
However, these FF's are generated via the AP evolution to the high scale
$\mu=\sqrt s$.
Thus, apart from the FF's of the heavy quarks $c,\bar c,b,\bar b$, also these
radiatively generated FF's contribute to the cross section.
All these contributions are properly included in the total result for
$(1/\sigma_{\rm tot})d\sigma/dx$ shown in Figs.~\ref{ee}a and b.
At LEP1 energies, the contribution from the first three quark flavors is
still negligible; it is concentrated at small $x$, and only amounts to a few
percent of the integrated cross section.
However, the contribution from the gluon FF, which appears at NLO in 
connection with three-parton final states, is numerically significant.
Motivated by the decomposition of $(1/\sigma_{\rm tot})d\sigma/dx$ in terms of
parton-level cross sections (see Eq.~(5) in Ref.~\cite{6}), in Figs.~\ref{ee}a
and b, we distributed this contribution over the $Z\to c\bar c$ and
$Z\to b\bar b$ channels in the proportion $e_c^2:e_b^2$, where $e_q$ is the
effective electroweak coupling of the quark $q$ to the $Z$ boson and the
photon including propagator adjustments.
This procedure approximately produces the quantities that should be compared
with the OPAL and ALEPH data.
We ignore the effect of electromagnetic initial-state radiation, which has not
been corrected for in the data.

As in Ref.~\cite{6}, we study the branching fractions for the transitions of
charm and bottom quarks to $D^{*\pm}$ mesons, defined by
\begin{equation}
\label{branchcut}
B_Q(\mu)=\int_{x_{\rm cut}}^1dxD_Q(x,\mu^2),
\end{equation}
where $Q=c,b$ and $x_{\rm cut} = 0.1$.
This allows us to test the consistency of our fits with information presented
in the experimental papers \cite{16,17}.
The contribution from the omitted region $0<x<x_{\rm cut}$ is close to zero.
For all four FF sets, we calculate $B_Q(\mu)$ at the respective threshold
$\mu=2m_Q$ and at the $Z$-boson resonance $\mu=M_Z$ and present the outcome in
Table~\ref{branching}.
As expected, the values of $B_Q(\mu)$ change very little under the evolution
from $2m_Q$ to $M_Z$, and they are very similar for $Q=c,b$.
The OPAL analysis \cite{16}, which is conceptually very different from ours,
yielded $B_c(M_Z)=0.222\pm0.014\pm0.014$ and
$B_b(M_Z)=0.173\pm0.016\pm0.012$, where the first (second) error is
statistical (systematic).
These values lie in the same ball park as our corresponding LO-O and NLO-O
results in Table~\ref{branching}.
The ALEPH paper \cite{17} does not explicitly quote values for these branching
fractions.
However, by combining other results given in Section~8 of Ref.~\cite{17}, we
infer that $B_c(M_z)=0.230{+0.070\atop-0.059}$, which nicely agrees with our
corresponding LO-A and NLO-A results in Table~\ref{branching}.

Another quantity of interest, which can directly be compared with experiment,
is the mean momentum fraction,
\begin{equation}
\label{average}
\langle x\rangle_Q(\mu)=\frac{1}{B_Q(\mu)}\int_{x_{\rm cut}}^1dx\,xD_Q(x,\mu),
\end{equation}
where $Q=c,b$.
In Table~\ref{xav}, we collect the values of $\langle x\rangle_Q(\mu)$ for
$Q=c,b$ evaluated at $\mu=2m_Q,M_Z$ with the four FF sets.
At fixed $\mu$, the differences between the OPAL and ALEPH sets and between LO
and NLO are marginal.
However, the AP evolution from $\mu=2m_Q$ to $\mu=M_Z$ leads to a significant 
reduction of $\langle x\rangle_Q(\mu)$, especially in the case of $Q=c$.
Our results for $\langle x\rangle_c(M_Z)$ should be compared with the 
experimental numbers reported by OPAL \cite{16} and ALEPH \cite{17},
\begin{eqnarray}
\langle x\rangle_c(M_Z)&=&0.515\pm0.002\pm0.009\qquad
({\rm OPAL}),
\nonumber\\
\langle x\rangle_c(M_Z)&=&0.4878\pm0.0046\pm0.0061\qquad
({\rm ALEPH}),
\label{avex}
\end{eqnarray}
respectively.
The central values are very close to those reported in the previous OPAL
\cite{4} and ALEPH \cite{3} publications, but the errors are now considerably 
smaller.
Comparing Eq.~(\ref{avex}) with Table~\ref{xav}, we conclude that our results
for $\langle x\rangle_c(M_Z)$ slightly undershoot the independent experimental
determinations in Refs.~\cite{16,17}, by about 5\% (11\%) at LO (NLO).
In order to assess the theoretical uncertainty of the results in
Table~\ref{xav} related to the choice of ansatz at the starting scale,
we recall that, in Ref.~\cite{6}, the LO-S (NLO-S) result for
$\langle x\rangle_c(M_Z)$ turned out to be 6\% (4\%) larger than the
LO-M (NLO-M) result.
At this point, we should also mention that the results in Eq.~(\ref{avex}) are
not directly extracted from the measured $x$ distributions, but from
calculations based on some Monte-Carlo model with parameters fitted to the
experimental data.
Whether this is the actual source of the difference remains unclear for the
time being.

Having constructed $D^{*\pm}$ FF's from LEP1 data, it is interesting to
quantitatively investigate whether they lead to a consistent description of
available data on $e^+e^-\to D^{*\pm}+X$ at other values of $\sqrt s$.
This would represent a direct test of the underlying scaling violation of 
fragmentation as implemented in the QCD-improved parton model via the
timelike AP equations.
To this end, we select the data from ARGUS \cite{15} at $\sqrt s=10.49$~GeV,
from HRS \cite{22} at $\sqrt s=29$~GeV, and from TASSO \cite{23} at
$\sqrt s=34.2$~GeV.
In Fig.~\ref{comp}a, we compare these data with our respective LO and NLO
predictions based on sets LO~O and NLO~O, respectively.
For reference, also the OPAL data \cite{16} are included.
The analogous comparison based on sets LO~A and NLO~A is shown in 
Fig.~\ref{comp}b.
To quantitatively assess these comparisons, we summarize the corresponding
values of $\chi_{\rm DF}^2$ in Table~\ref{chi}.
We observe that the scaling violation encoded in the experimental data is 
faithfully described by our theoretical predictions.
As expected on general grounds, the $\chi_{\rm DF}^2$ values tend to be lower
for the NLO analyses.
Within the framework of the QCD-improved parton model, the OPAL data appear to
be more consistent with the ARGUS data than the ALEPH data, while the ALEPH
data seem agree better with the HRS data than the OPAL data do.
However, we should bear in mind that, in contrast to the LEP1 data, the ARGUS,
HRS, and TASSO data are presented in the form $sd\sigma/dx$ and thus suffer
from an additional normalization uncertainty.

The successful comparisons in Figs.~\ref{comp}a and b and Table~\ref{chi}
reassure us that our $D^{*\pm}$ FF's, although constructed at $\sqrt s=M_Z$,
also lead to useful descriptions of $D^{*\pm}$ fragmentation at other scales.
In the next section, we exploit this property together with the universality
of fragmentation to make predictions for inclusive $D^{*\pm}$ photoproduction
at HERA.

\boldmath
\section{$D^{*\pm}$ Production in Low-$Q^2$ $ep$ Collisions}
\unboldmath

In this section, we compare our NLO predictions for the cross section of
inclusive $D^{*\pm}$ photoproduction in $ep$ scattering at HERA with the 1996
data from the ZEUS collaboration, which were presented by Y.~Eisenberg at the
1997 International Europhysics Conference on High Energy Physics \cite{18}.
We emphasize that these data are still preliminary.

The present HERA conditions are such that $E_p=820$~GeV protons collide with
$E_e= 27.5$~GeV positrons in the laboratory frame.
The rapidity is taken to be positive in the proton flight direction.
The quasi-real photon spectrum is described in the Weizs\"acker-Williams
approximation by Eq.~(5) of Ref.~\cite{14}. 
This spectrum depends on the photon-energy fraction, $x=E_\gamma/E_e$, and the
maximum photon-virtuality, $Q_{\rm max}^2$.
In the ZEUS experiment \cite{2,18}, where the final-state electron is not
detected, $Q_{\rm max}^2=4$~GeV$^2$ and $0.147<x<0.869$, which corresponds to
$\gamma p$ CM energies in the range 115~GeV${}<W<280$~GeV.
We adopt all these kinematic conditions in our analysis.
We work at NLO in the $\overline{\rm MS}$ scheme with $n_f=4$ flavors.
For the proton and photon PDF's we use set CTEQ4M \cite{24} with 
$\Lambda_{\overline{\rm MS}}^{(4)}=296$~MeV and set GRV~HO \cite{25} converted
to the $\overline{\rm MS}$ factorization scheme, respectively.
We evaluate $\alpha_s(\mu^2)$ from the two-loop formula with this value of
$\Lambda_{\overline{\rm MS}}^{(4)}$.
The $\Lambda_{\overline{\rm MS}}^{(4)}$ values implemented in the photon PDF's
and the $D^{*\pm}$ FF's are 200~MeV and 352~MeV, which corresponds to the value
$\Lambda_{\overline{\rm MS}}^{(5)}=227$~MeV quoted in Sec.~2, respectively.
We identify the factorization scales associated with the proton, photon, and
$D^{*\pm}$ mesons, and collectively denote them by $M_f$. 
We choose the renormalization and factorization scales to be
$\mu=m_T$ and $M_f=2m_T$, respectively, where $m_T=\sqrt{p_T^2+m_c^2}$ is the
$D^{*\pm}$ transverse mass.
Whenever we present LO results, these are consistently computed using set
CTEQ4L \cite{24} of proton PDF's, set GRV~LO \cite{25} of photon PDF's, the LO
versions of our two sets of $D^{*\pm}$ FF's, the one-loop formula for
$\alpha_s$ with $\Lambda_{\overline{\rm MS}}^{(4)}=236$~MeV \cite{24}, and the
LO hard-scattering cross sections.

The photoproduction cross section is a superposition of the direct- and
resolved-photon contributions.
In our NLO analysis, the resolved-photon contribution is larger than the
direct one for moderate $p_T$. 
This statement depends, however, on the factorization scheme;
only the sum of both contributions is a physical observable and can be
compared with experimental data. 
The bulk of the resolved-photon cross section is due to the charm content of
the photon \cite{14}.

The ZEUS data \cite{18} come as three distributions:
(i) the $p_T$ distribution $d\sigma/dp_T$ integrated over
$-1.5<y_{\rm lab}<1$ and 115~GeV${}<W<280$~GeV;
(ii) the $y_{\rm lab}$ distribution $d\sigma/dy_{\rm lab}$ integrated over
4~GeV${}<p_T<12$~GeV and 115~GeV${}<W<280$~GeV; and
(iii) the $W$ distribution $d\sigma/dW$ integrated over
4~GeV${}<p_T<12$~GeV and $-1.5<y_{\rm lab}<1$.
In Figs.~\ref{zeus}a--c, we compare the measured distributions (i)--(iii) with
our LO and NLO predictions based on FF sets LO~O, LO~A, NLO~O, and NLO~A.
In all three cases, the LO-O (NLO-O) result slightly exceeds the LO-A (NLO-A)
result, but the differences are still small compared to the experimental 
uncertainty of the ZEUS data.
This means that the details of charm fragmentation are sufficiently well
constrained by the LEP1 data, and that the HERA data are not yet precise 
enough to resolve the presently existing difference between the OPAL \cite{16}
and ALEPH \cite{17} data.
In Figs.~\ref{zeus}a--c, the shifts due to use of different experimental input
(OPAL versus ALEPH) tend to be less significant than the effects due to the
inclusion of higher-order corrections (LO versus NLO).
However, the differences between the LO and NLO results are still considerably
smaller than the errors on the ZEUS data, and they are approximately equal for
the OPAL and ALEPH sets.
The similarity of the LO and NLO results indicates good perturbative
stability.
Apparently, this is a special virtue of the pure $\overline{\rm MS}$
factorization scheme with strictly massless quark flavors, which we are using
here.
By contrast, we encountered a sizeable gap between the LO and NLO results in
the massive subtraction scheme \cite{6}.

In Fig.~\ref{zeus}a, the NLO distributions fall off slightly less strongly
with increasing $p_T$ than the LO ones, and thus agree better with the data at
large $p_T$.
The overall agreement with the data is very satisfactory, even at small $p_T$,
where the massless approach ceases to be valid.
The theoretical predictions somewhat undershoot the measurement in the highest
$p_T$ bin, but the experimental error is still rather sizeable there.
As for Fig.~\ref{zeus}b, the effect of the higher-order corrections and the
influence of the experimental input is most pronounced in the backward
direction, where all theoretical predictions agree with the experimental data
within one standard deviation.
At $y_{\rm lab}=-1.5$, the NLO-O result exceeds the LO-O one by 25\% and the
NLO-A one by 15\%.
On the other hand, in the forward direction, at $y_{\rm lab}>0$, the ZEUS data
points significantly overshoot the theoretical predictions.
There, the experimental errors are largest, while the theoretical uncertainty 
is relatively small.
In fact, both the NLO corrections and the uncertainty due to spread of the LEP1
data are negligibly small there.
Furthermore, we know from Ref.~\cite{6} that the specific choice of ansatz for
the charm FF's at the starting scale is numerically irrelevant.
Changes in $\Lambda_{\overline{\rm MS}}$ or $\mu_0$ are expected to be 
compensated through the fit by appropriate shifts in the input parameters of
Eqs.~(\ref{peterson}) and (\ref{standard}), and do not represent an
appreciable theoretical error source either.
The variation in cross section due to the use of different proton PDF's is
just of order 10\%, as long as up-to-date sets are considered \cite{14}.
The most substantial source of theoretical uncertainty is related to the
photon PDF's.
Among the different sets used in Ref.~\cite{6}, GRV~HO \cite{25} gave the
smallest cross section at $y_{\rm lab}>0$; the variation was found to be about
40\%.
An increase of this size would render the theoretical predictions compatible 
with the ZEUS data point at $y_{\rm lab}=0.75$, while it would be far to small
to explain the data point at $y_{\rm lab}=1.25$, which represents an
unexpected rise in cross section.
It remains to be seen if such a rise will also be observed by the H1 
collaboration.
Looking at Fig.~\ref{zeus}c, we observe that, within the rather large
experimental errors, all four theoretical predictions agree reasonably well
with the data.
The NLO-O result leads to the best description of the data.
Only the data point at $W=252.5$~GeV is significantly above the theoretical
expectation.

In the ZEUS paper \cite{18}, the $y_{\rm lab}$ and $W$ distributions were also
presented for a minimum-$p_T$ cut of 3~GeV.
We performed similar comparisons with these data, too, but do not display them
here, since our massless-charm and -bottom scheme is more appropriate for high
values of $p_T$.
In the case of the $y_{\rm lab}$ distribution, the general picture is very 
similar to Fig.~\ref{zeus}b, except that the data point at $y_{\rm lab}=0.75$
is then consistent with the theoretical expectation.
Similarly, in the case of the $W$ distribution, the data point at
$W=252.5$~GeV then agrees considerably better with the theoretical prediction.

\section{Conclusions}

The OPAL and ALEPH collaborations presented measurements of the fractional
energy spectrum of inclusive $D^{*\pm}$ production in $Z$-boson decays based
on their entire LEP1 data samples \cite{16,17}.
Apart from the full cross section of $D^{*\pm}$ production, they also
determined the contribution arising from $Z\to b\bar b$ decays.
This enabled us to update our LO and NLO fits of the $D^{*\pm}$ FF's \cite{6},
which were based on previous OPAL and ALEPH analyses with considerably lower
statistics \cite{3,4}.

At the same time, we also incorporated a conceptual modification of our 
theoretical framework, so as to eliminate two minor weaknesses of our
previous approach \cite{6}:
(i) At NLO, the differential cross section $d\sigma/dx$ of
$e^+e^-\to D^{*\pm}+X$ turned negative for $x\agt0.9$, so that the 
experimental data points in this region needed to be excluded from the fits;
this problem was also encountered in Ref.~\cite{5}.
(ii) The LO and NLO results for the cross section of inclusive $D^{*\pm}$
photoproduction in $ep$ collisions substantially differed from each other,
indicating that, in this particular scheme, higher orders beyond present 
control are likely to be significant.
In Ref.~\cite{6}, the factorization of the final-state collinear singularities
associated with the charm and bottom quarks was performed in a scheme which
differs from the pure $\overline{\rm MS}$ scheme with five massless quark
flavors by the subtraction of certain finite functions from the parton-level
cross sections.
These functions were chosen in such a way that the resulting change of scheme
is equivalent to the use of the perturbative FF's of Ref.~\cite{8}.
Since these perturbative FF's describe the production of a massive quark $Q$
rather than a heavy hadron $H$, they need to be complemented by a
nonperturbative component which accounts for the $Q\to H$ fragmentation 
process.
In particular, this is requisite to make sure that the $Q\to H$ branching
ratio and the average $H$ to $Q$ longitudinal-momentum fraction take their
measured values, which are indeed smaller than unity.
Since the charm quark is only moderately heavy, the effect of this
nonperturbative component is rather dramatic, and its omission renders the
theoretical description completely inadequate, as was demonstrated in
Ref.~\cite{14}.
Clearly, the convolution of a perturbative FF with a nonperturbative ansatz
can in turn be considered as a nonperturbative FF.
This consideration brings us back to the pure $\overline{\rm MS}$ scheme with
five massless flavors, from which we started off.
In this paper, we advocated the application of this scheme to charmed-meson
production at high $\sqrt s$ in $e^+e^-$ annihilation and at large $p_T$ in
$ep$ scattering or similar types of collision.
As a matter of principle, the question whether this approach is justified in 
the present case or not can only be answered by nature itself.
We found that the pure $\overline{\rm MS}$ scheme is clearly favoured by a
wealth of experimental data on $D^{*\pm}$ production \cite{15,16,17,18,22,23}.
In fact, both drawbacks mentioned above are nicely avoided.
Furthermore, comparing Figs.~\ref{comp}a and b with Fig.~3 in Ref.~\cite{6},
we observe that the general description of the $e^+e^-$ data from DORIS
\cite{15}, PEP \cite{22}, and PETRA \cite{23} is improved, especially in the
upper $x$ range.

It is important to bear in mind that the fit results for the input parameters
in Eqs.~(\ref{peterson}) and (\ref{standard}), including the value of
Peterson's $\epsilon_c$, are highly scheme dependent at NLO, and must not be
na\"\i vely compared disregarding the theoretical framework which they refer
to.
If we compare the values of $N$ and $\epsilon_c$ for the charm FF of set NLO~M
previously obtained in the massive subtraction scheme \cite{6} with those of
set NLO~O appropriate to the pure $\overline{\rm MS}$ scheme, we find dramatic
differences: $N=0.0677$ and $\epsilon_c=0.0204$ for set NLO~M versus $N=0.267$
and $\epsilon_c=0.116$ for set NLO~O.
On the other hand, the respective LO results are almost identical ($N=0.202$
and $\epsilon_c=0.0856$ for set LO~M versus $N=0.223$ and $\epsilon_c=0.0851$
for set LO~O), which indicates that the underlying data (from Refs.~\cite{3,4}
for set M and from Ref.~\cite{16} for set O) are consistent with each other.

Finally, we should caution the reader that the massless-quark approximation
used here and the resulting FF's are only appropriate for processes of
$D^{*\pm}$ production which are characterized by an energy scale that is large
against the charm-quark mass.
In particular, this approach should not be expected to yield meaningful
predictions for the photoproduction of small-$p_T$ $D^{*\pm}$ mesons at HERA.
Therefore, in order to substantiate the comparison of LEP1 and HERA data in
the framework of the QCD-improved parton model endowed with nonperturbative
FF's, it would be very desirable if the statistics of the HERA data was
increased in the large-$p_T$ bins.

\bigskip
\centerline{\bf ACKNOWLEDGMENTS}
\smallskip\noindent
We are grateful to Ties Behnke and Stefan S\"oldner-Rembold of OPAL, to Paul
Colas of ALEPH, and to Carsten Coldewey of ZEUS for making available to us the
respective data on $D^{*\pm}$ production \cite{16,17,18} in numerical form.
JB and GK thank the Theory Group of the Werner-Heisenberg-Institut for the
hospitality extended to them during visits when this paper was prepared.

\newpage

\begin{table}
\centerline{\bf TABLE CAPTIONS}
\bigskip

\caption{Fit parameters for the charm- and bottom-quark FF's of sets LO~O,
LO~A, NLO~O, and NLO~A.
The corresponding starting scales are $\mu_0=2m_c=3$~GeV and 
$\mu_0=2m_b=10$~GeV, respectively.
All other FF's are taken to be zero at $\mu_0=2m_c$
\protect\label{pars}}

\caption{$\chi^2$ per degree of freedom pertaining to the LO and NLO fits to
the OPAL \protect\cite{16} and ALEPH \protect\cite{17} data.
In each case, $\chi_{\rm DF}^2$ is calculated for the $Z\to b\bar b$ sample,
the full sample, and the combination of both.
\protect\label{fit}}

\caption{Branching fractions of charm and bottom quarks into $D^{*\pm}$ mesons
at the respective starting scales and at $\mu=M_Z$ evaluated from 
Eq.~(\protect\ref{branchcut}) with sets LO~O, LO~A, NLO~O, and NLO~A.
\protect\label{branching}}

\caption{Average momentum fractions of $D^{*\pm}$ mesons produced through 
charm- and bottom-quark fragmentation at the respective starting scales and at
$\mu=M_Z$ evaluated from Eq.~(\protect\ref{average}) with sets LO~O, LO~A,
NLO~O, and NLO~A.
\protect\label{xav}}

\caption{$\chi^2$ per degree of freedom evaluated with sets LO~O, LO~A, NLO~O,
and NLO~A for the ARGUS \protect\cite{15}, HRS \protect\cite{22}, and TASSO
\protect\cite{23} data.
TASSO1 (TASSO~2) refers to the $D^0\to K^-\pi^+\pi^+\pi^-$ ($D^0\to K^-\pi^+$)
channel.
\protect\label{chi}}

\end{table}

\newpage

\begin{figure}
\centerline{\bf FIGURE CAPTIONS}
\bigskip

\caption{(a) The cross sections of inclusive $D^{*\pm}$ production in $e^+e^-$
annihilation evaluated with sets LO~O and NLO~O are compared with the OPAL
data \protect\cite{16}.
The three pairs of curves correspond to the $Z\to c\bar c$, $Z\to b\bar b$,
and full samples.
(b) The same for sets LO~A and NLO~A and the ALEPH data \protect\cite{17}.
\protect\label{ee}\hskip8cm}

\caption{(a) The ARGUS \protect\cite{15}, HRS \protect\cite{22}, TASSO
\protect\cite{23}, and OPAL \protect\cite{16} data on inclusive $D^{*\pm}$
production in $e^+e^-$ annihilation are compared with the predictions based on
sets LO~O and NLO~O.
(b) The same for sets LO~A and NLO~A and the ALEPH data \protect\cite{17}.
For separation, the data have been rescaled by powers of 10.
In the case of TASSO, the open triangles refer to the
$D^0\to K^-\pi^+\pi^+\pi^-$ channel (TASSO~1) and the solid triangles to the
$D^0\to K^-\pi^+$ channel (TASSO~2).
In the cases of OPAL and ALEPH, we consider the dimensionless quantity
$(1/\sigma_{\rm tot})d\sigma/dx$.
\protect\label{comp}}

\caption{The predictions of inclusive $D^{*\pm}$ photoproduction in $ep$
collisions based on sets LO~O, LO~A, NLO~O, and NLO~A are compared with the
ZEUS data \protect\cite{18}.
We consider (a) the $p_T$ distribution $d\sigma/dp_T$ integrated over
$-1.5<y_{\rm lab}<1$ and 115~GeV${}<W<280$~GeV,
(b) the $y_{\rm lab}$ distribution $d\sigma/dy_{\rm lab}$ integrated over
4~GeV${}<p_T<12$~GeV and 115~GeV${}<W<280$~GeV, and
(c) the $W$ distribution $d\sigma/dW$ integrated over
4~GeV${}<p_T<12$~GeV and $-1.5<y_{\rm lab}<1$.
\protect\label{zeus}}

\end{figure}

\newpage

\begin{table}
\begin{center}
\begin{tabular}{|c|c|c||c|c|c|c|}
\hline
collab. & set & flavour & $N$   & $\alpha$ & $\beta$ & $\epsilon_c$ \\
\hline
\hline
      & LO  & $c$ & 0.223 & --   & --   & 0.0851 \\
\cline{3-7}
OPAL  &     & $b$ & 84.0  & 2.69 & 5.13 & -- \\
\cline{2-7}
      & NLO & $c$ & 0.267 & --   & --   & 0.116 \\
\cline{3-7}
      &     & $b$ & 18.9  & 1.71 & 4.02 & -- \\
\hline
\hline
      & LO  & $c$ & 0.385 & --   & --   & 0.144 \\
\cline{3-7}
ALEPH &     & $b$ & 196   & 2.98 & 6.21 & -- \\
\cline{2-7} 
      & NLO & $c$ & 0.444 & --   & --   & 0.185 \\
\cline{3-7}
      &     & $b$ & 86.4  & 2.41 & 5.96 & -- \\
\hline 
\end{tabular}

\smallskip
{\bf Table~1}

\vspace{0.5cm}

\begin{tabular}{|c||c|c|c||c|c|c|}
\hline
set  & \multicolumn{3}{|c||}{OPAL}& \multicolumn{3}{|c|}{ALEPH}\\
     & ave. & $b$  & sum & ave. & $b$  & sum \\
\hline
\hline
LO   & 1.11 & 1.11 & 1.11  & 1.53  & 1.65 & 1.42 \\
\hline
NLO  & 0.92 & 0.97 & 0.88  & 1.39  & 1.60 & 1.22 \\
\hline
\end{tabular}

\smallskip
{\bf Table~2}

\vspace{0.5cm}

\begin{tabular}{|c|c||c|c|c|c|}
\hline
collab. & set & $B_c(2m_c)$ & $B_c(M_Z)$ & $B_b(2m_b)$ & $B_b(M_Z)$ \\
\hline
\hline 
OPAL  & LO   & 0.273 & 0.256 & 0.246 & 0.232 \\
\cline{2-6}
      & NLO  & 0.255 & 0.238 & 0.238 & 0.220 \\
\hline
\hline
ALEPH & LO   & 0.308 & 0.287 & 0.213 & 0.200 \\
\cline{2-6}
      & NLO  & 0.288 & 0.265 & 0.201 & 0.186 \\
\hline
\end{tabular}

\smallskip
{\bf Table~3}

\vspace{0.5cm}

\begin{tabular}{|c|c||c|c|c|c|}
\hline
collab. & set & $\langle x\rangle_c(2m_c)$ & $\langle x\rangle_c(M_Z)$ &
     $\langle x\rangle_b(2m_b)$ & $\langle x\rangle_b(M_Z)$ \\
\hline 
\hline
OPAL  & LO  & 0.644 & 0.490 & 0.391 & 0.343 \\
\cline{2-6}
      & NLO & 0.617 & 0.464 & 0.362 & 0.320 \\
\hline
\hline
ALEPH & LO  & 0.598 & 0.459 & 0.360 & 0.317 \\
\cline{2-6} 
      & NLO & 0.576 & 0.437 & 0.337 & 0.298 \\
\hline
\end{tabular}

\smallskip
{\bf Table~4}

\vspace{0.5cm}

\begin{tabular}{|c|c||c|c|c|c|}
\hline
collab. & set & ARGUS & HRS & TASSO1 & TASSO2 \\
\hline
\hline 
OPAL  & LO   & 1.86 & 2.01 & 1.20 & 1.84 \\
\cline{2-6}
      & NLO  & 0.86 & 2.02 & 0.95 & 1.54 \\
\hline
\hline
ALEPH & LO   & 4.67 & 1.00 & 0.76 & 2.35 \\
\cline{2-6}
      & NLO  & 3.43 & 0.96 & 0.59 & 2.09 \\
\hline
\end{tabular}

\smallskip
{\bf Table~5}

\end {center}
\end{table}

\newpage
\begin{figure}[ht]
\epsfig{figure=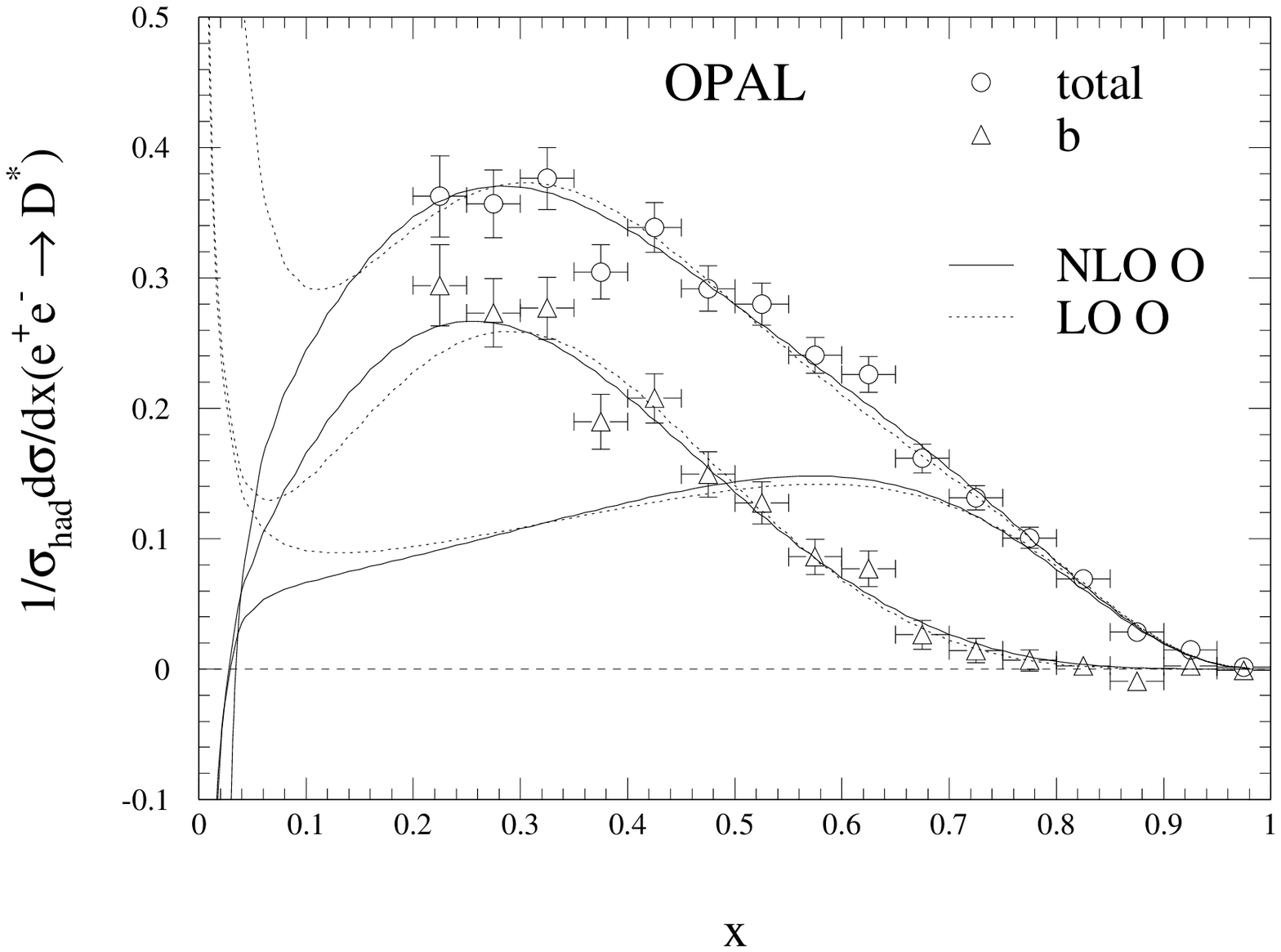,width=\textwidth}
\centerline{\Large\bf Fig.~1a}
\end{figure}

\newpage
\begin{figure}[ht]
\epsfig{figure=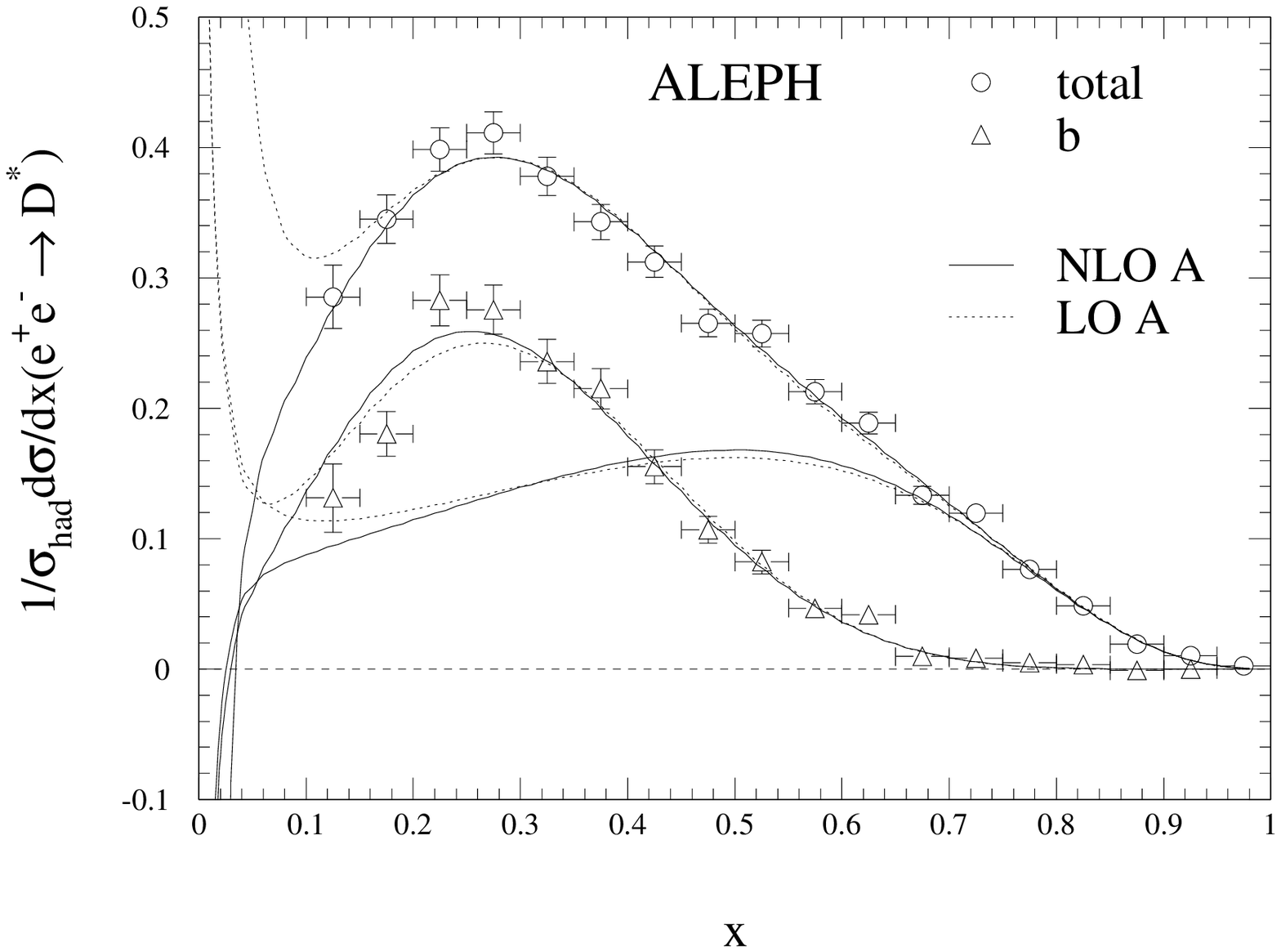,width=\textwidth}
\centerline{\Large\bf Fig.~1b}
\end{figure}

\newpage
\begin{figure}[ht]
\epsfig{figure=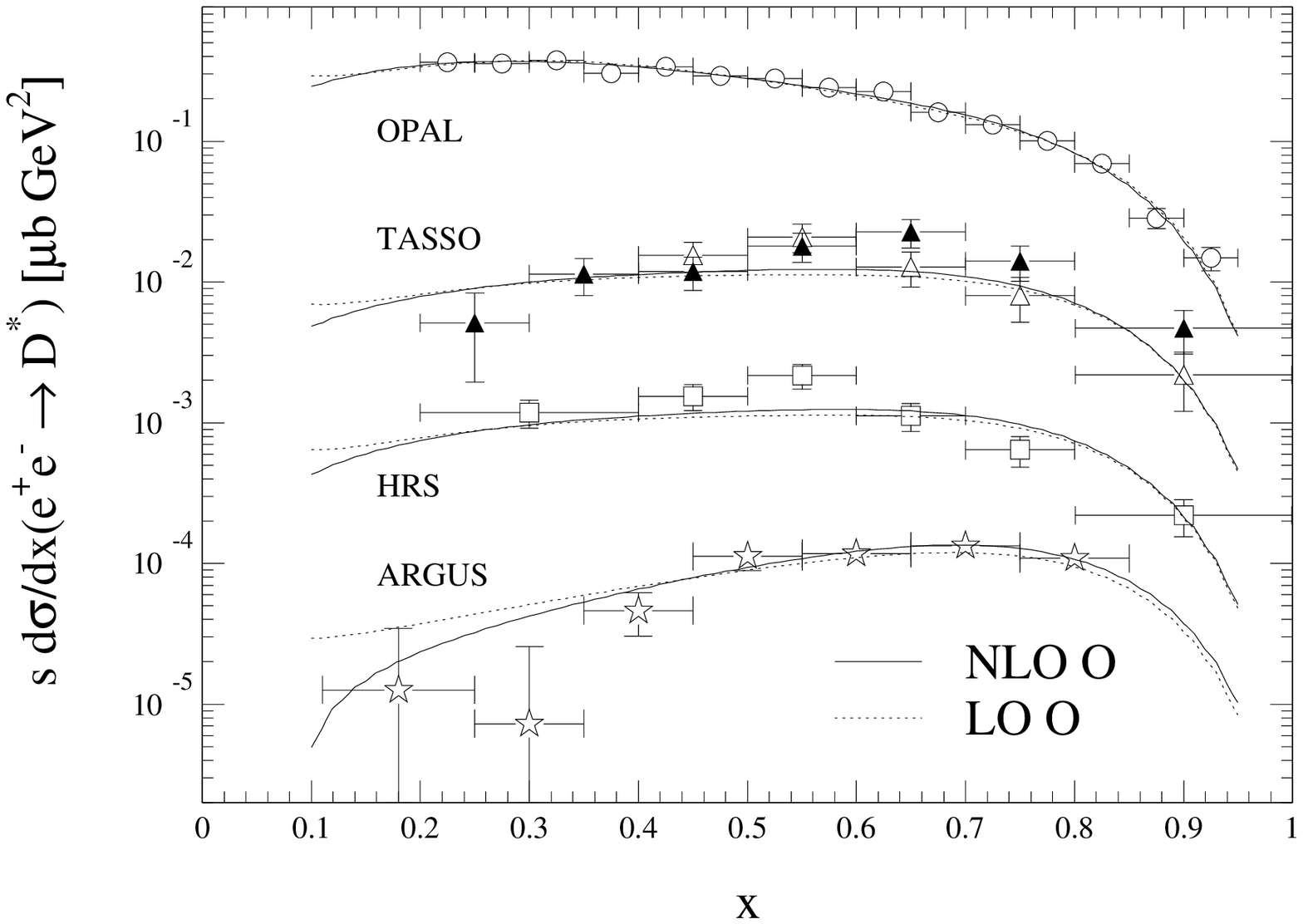,width=\textwidth}
\centerline{\Large\bf Fig.~2a}
\end{figure}

\newpage
\begin{figure}[ht]
\epsfig{figure=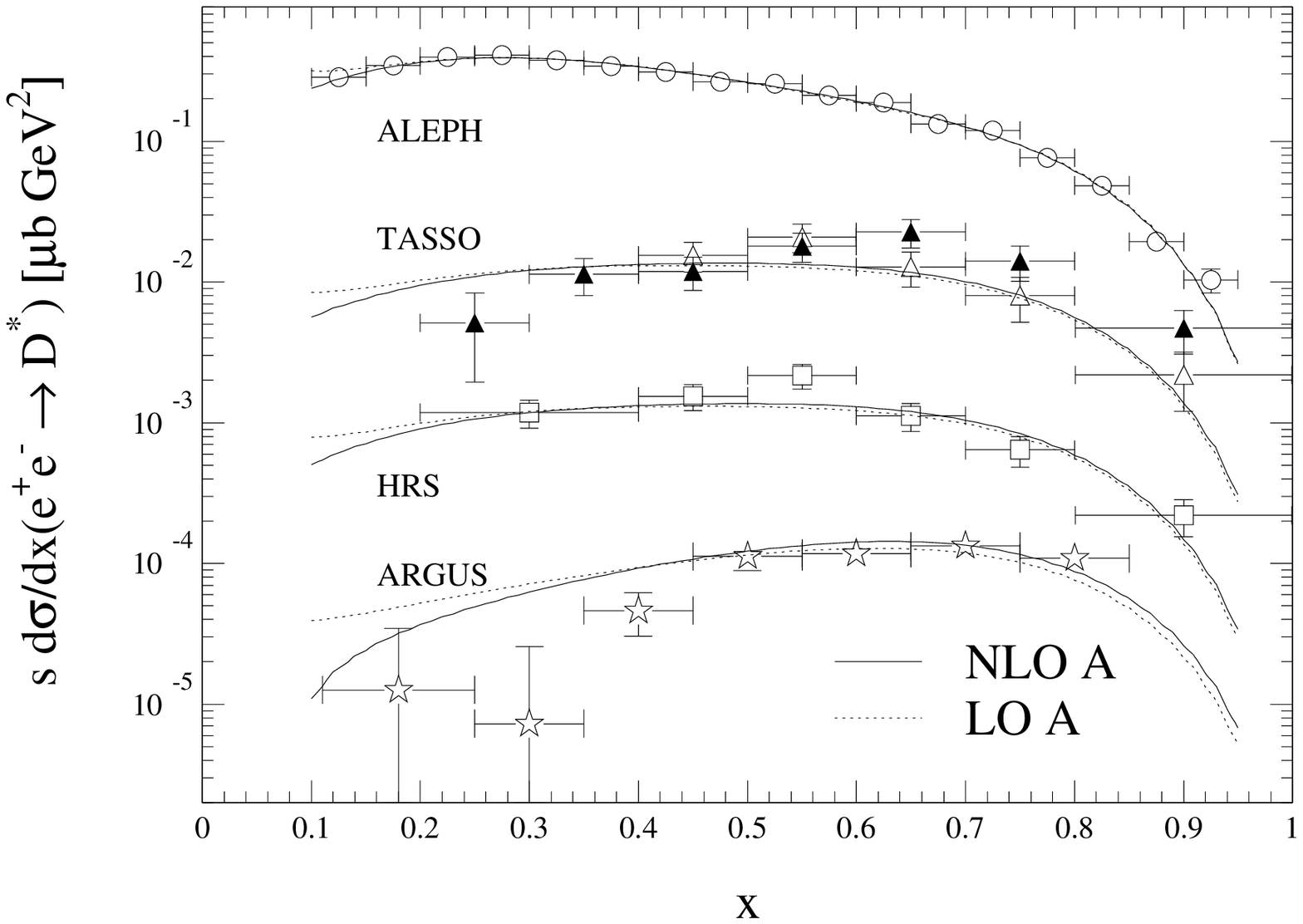,width=\textwidth}
\centerline{\Large\bf Fig.~2b}
\end{figure}

\newpage
\begin{figure}[ht]
\epsfig{figure=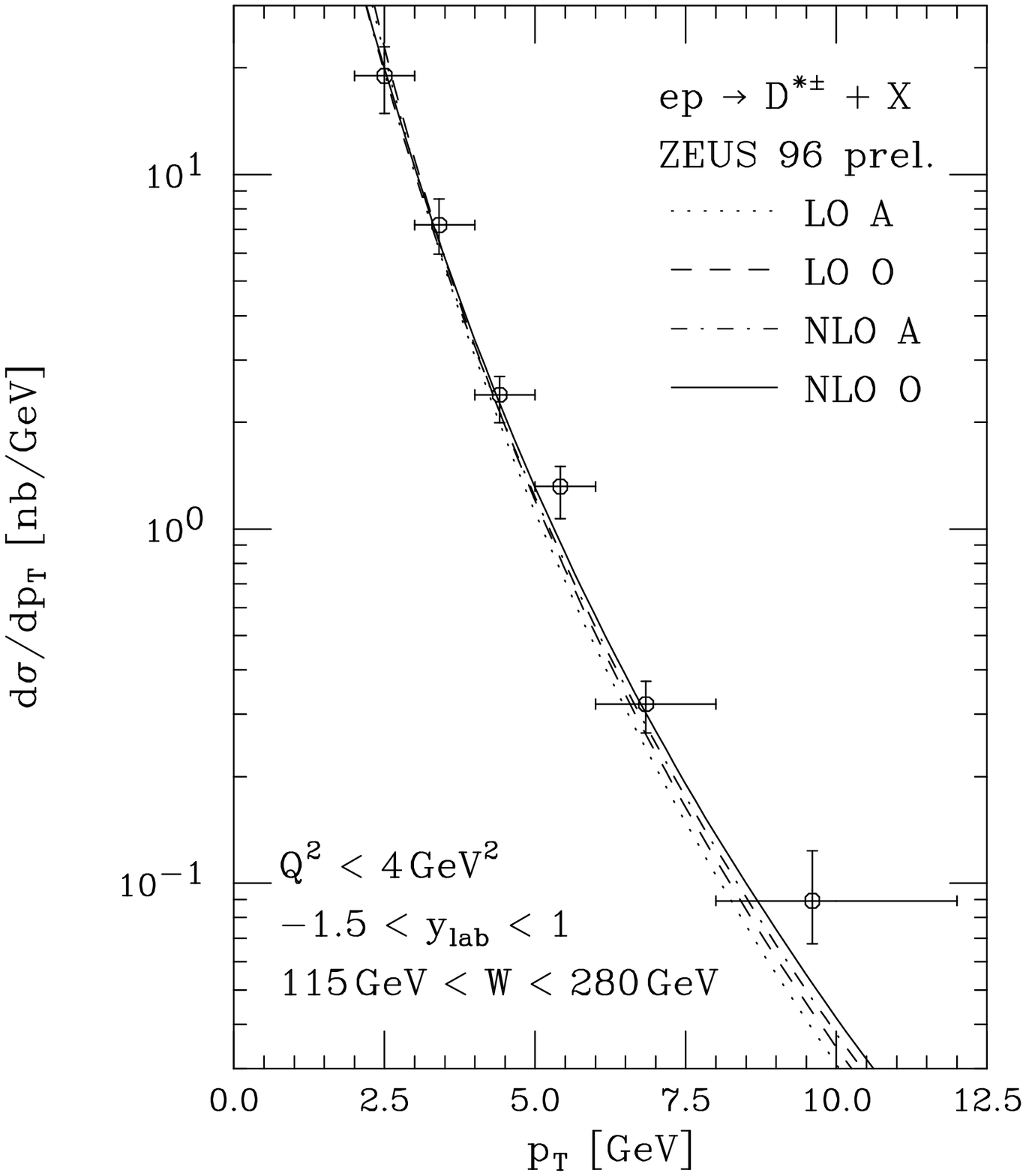,width=\textwidth}
\centerline{\Large\bf Fig.~3a}
\end{figure}

\newpage
\begin{figure}[ht]
\epsfig{figure=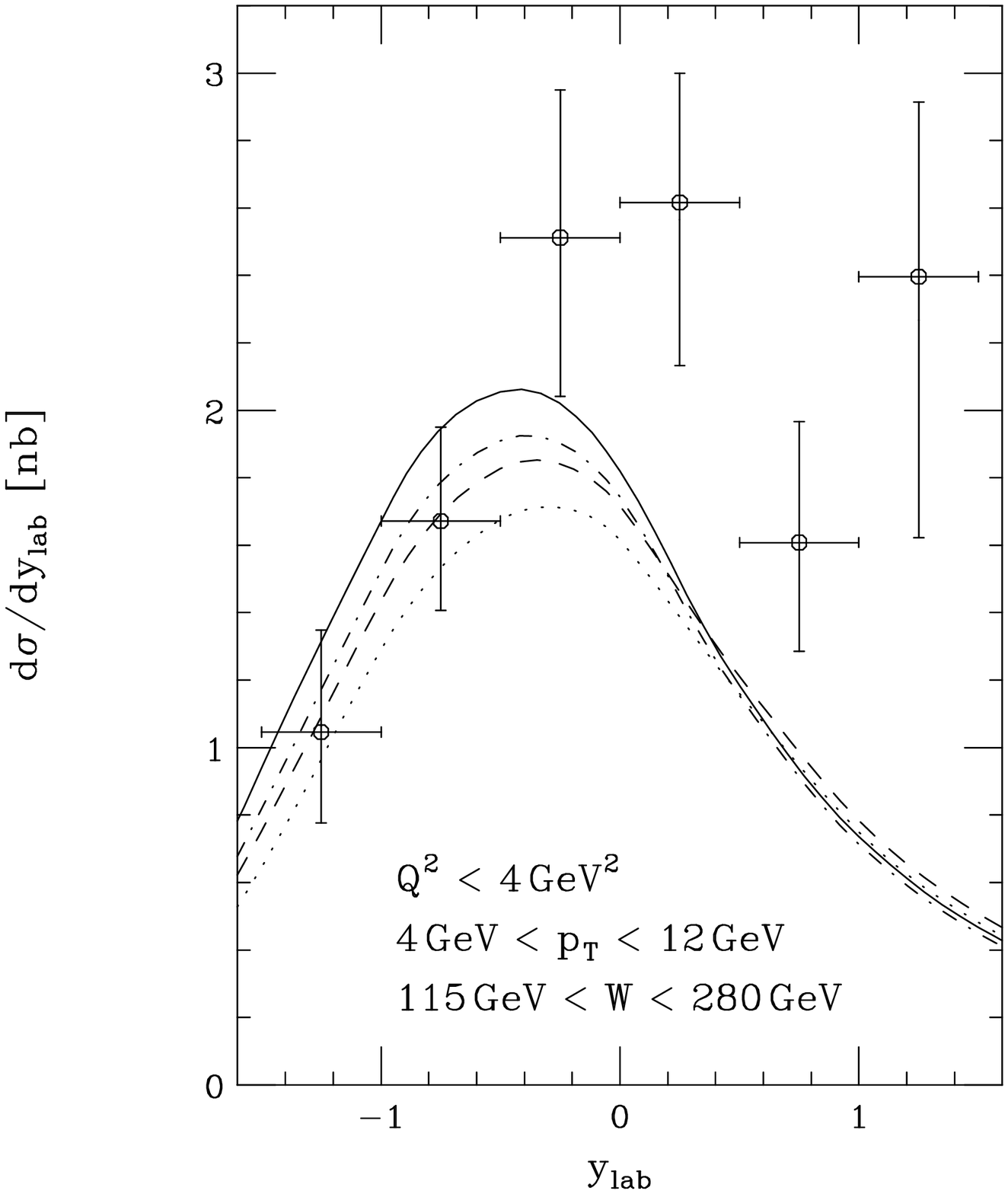,width=\textwidth}
\centerline{\Large\bf Fig.~3b}
\end{figure}

\newpage
\begin{figure}[ht]
\epsfig{figure=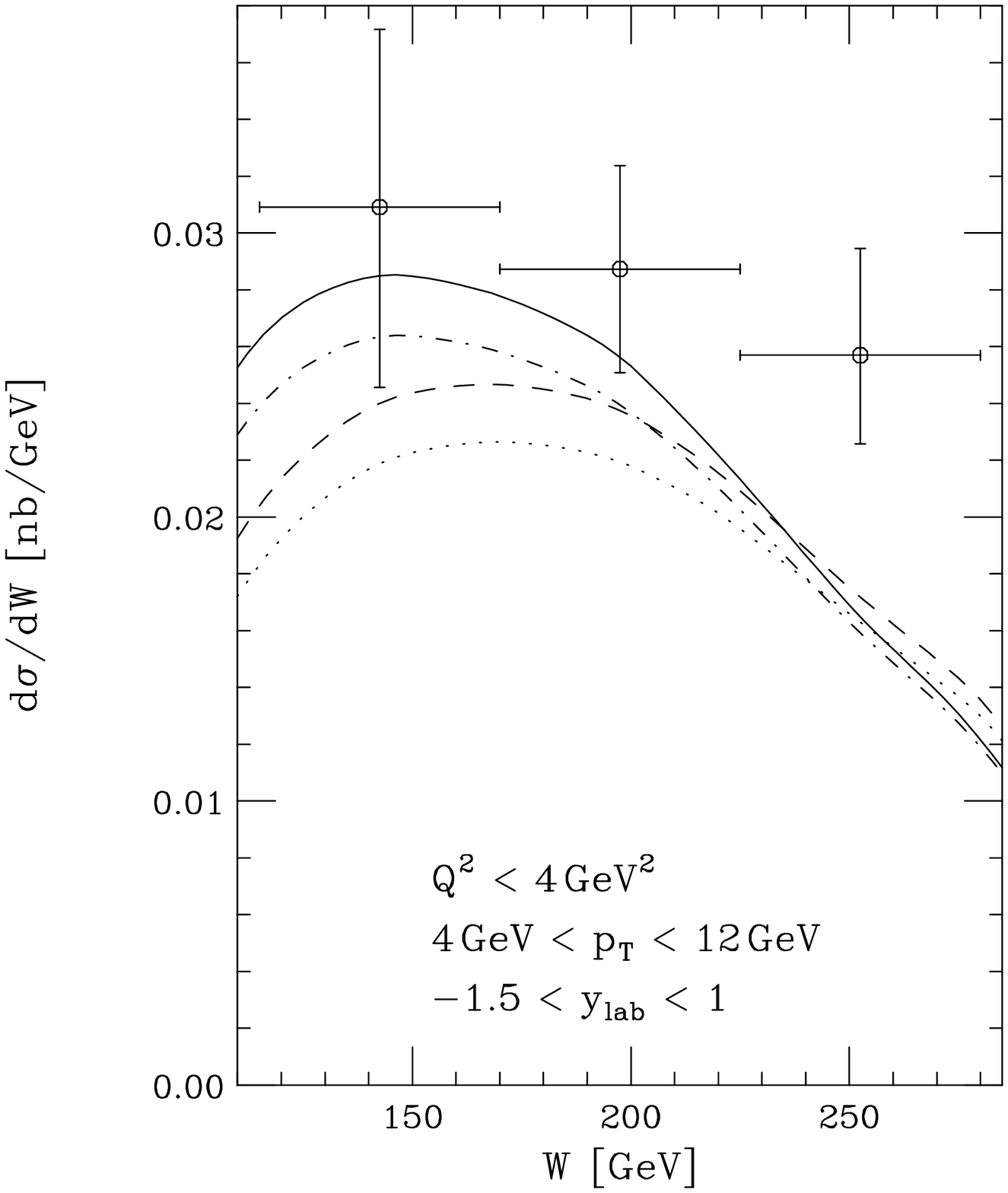,width=\textwidth}
\centerline{\Large\bf Fig.~3c}
\end{figure}

\end{document}